# Opportunities for Neutrino Physics at the Spallation Neutron Source (SNS)

**Yu Efremenko[1,2] and W R Hix[2,1]**
[1]University of Tennessee, Knoxville TN 37919, USA
[2]Oak Ridge National Laboratory, Oak Ridge TN 37981, USA

E-mail: yefremen@utk.edu

**Abstract**. In this paper we discuss opportunities for a neutrino program at the Spallation Neutrons Source (SNS) being commissioning at ORNL. Possible investigations can include study of neutrino-nuclear cross sections in the energy rage important for supernova dynamics and neutrino nucleosynthesis, search for neutrino-nucleus coherent scattering, and various tests of the standard model of electro-weak interactions.

## 1. Introduction

It seems that only yesterday we gathered together here at Columbia for the first Carolina Neutrino Symposium on Neutrino Physics. To my great astonishment I realized it was already eight years ago. However by looking back we can see that enormous progress has been achieved in the field of neutrino science since that first meeting. Eight years ago we did not know which region of mixing parameters (SMA. LMA, LOW, Vac) [1] would explain the solar neutrino deficit. We did not know whether this deficit is due to neutrino oscillations or some other even more exotic phenomena, like neutrinos decay [2], or due to the some other effects [3]. Hints of neutrino oscillation of atmospheric neutrinos had not been confirmed in accelerator experiments. Double beta decay collaborations were just starting to think about experiments with sensitive masses of hundreds of kilograms. Eight years ago, very few considered that neutrinos can be used as a tool to study the Earth interior [4] or for non-proliferation [5]. This impressive success in neutrino physics lead to increasing interest to this field. It is growing area of research and more and more young students are entering the field of neutrino physics. In this article we are going to discuss opportunities for a comprehensive neutrino experimental program that could start in ORNL at SNS.

## 2. Neutrino Production at SNS

Let us begin by describing the mechanism of neutrino production at the SNS. The Spallation Neutron Source, currently under commissioning at ORNL, will be the world's premier facility for neutron-scattering research, producing pulsed neutron beams with intensities an order of magnitude larger than any currently operating facility. When full beam power is reached in 2009, $10^{14}$ one GeV protons will bombard a liquid mercury target in 700 ns wide bursts with a frequency of 60 Hz. Neutrons produced by spallation reactions with the mercury will thermalize in hydrogenous and helium moderators surrounding the target and be delivered to neutron scattering instruments in the SNS experimental hall.

As a by-product, the SNS will also provide the world's most intense pulsed source of neutrinos, in the energy regime of interest for nuclear astrophysics. Interactions of the proton beam in the mercury target will produce π-mesons in addition to neutrons. These will stop inside the dense mercury target and their subsequent decay chain will produce neutrinos with a flux of ~$2\times10^7$ ν/cm$^2$/s for three neutrino flavors at 20 m from the spallation target [6].

The time structure of the SNS beam is particularly advantageous for neutrino studies. Time correlations between candidate events and the SNS proton beam pulse will greatly reduce background rates and may provide sensitivity to neutral current events. As shown in the left panel of Figure 1, all neutrinos will be produced within several microseconds of the 60 Hz proton beam pulses. As a result, background events resulting from cosmic rays will be suppressed by a factor of ~2000. At the beginning of the beam spill the neutrino flux is dominated by muon neutrinos resulting from pion decay, making it possible to isolate pure neutral-current events. This time structure provides a great advantage over a continuous-beam facility such as the Los Alamos Neutron Science Centre (LANSCE), where the LSND [7] experiment was located.

The energy spectra of SNS neutrinos are shown in the right hand panel of Figure 1. These spectra are known with high accuracy because nearly all neutrinos originate from decay-at-rest processes in which the kinematics is well defined. The decay of stopped pions produces mono-energetic muon neutrinos at 30 MeV. The subsequent 3-body muon decay produces a spectrum of electron neutrinos and muon antineutrinos with energies up to 52.6 MeV. Decay chains for negative pions and muons are strongly suppressed by their quick absorption in the bulk of mercury target.

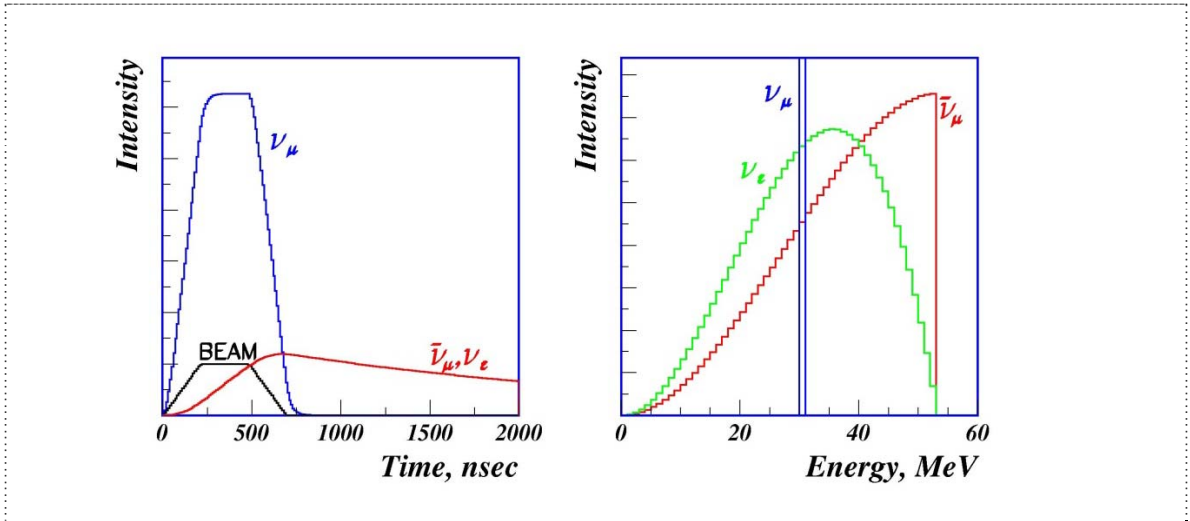

**Figure 1.** Time and energy distributions for different neutrino flavors produced at the SNS.

### 3. Neutrino Nucleus Cross Sections

As it been mention at many occasions [8], [9] that neutrino energy spectra from SNS nicely match the neutrino spectra produced during core collapse supernovae (SN) explosions. This gives us a unique opportunity to study neutrino interaction in this important energy range, which will improve our understanding of SN dynamics and help us to design and calibrate the response of supernova neutrino detectors.

Core collapse supernovae, which are among the most energetic explosions in our universe, releasing $10^{46}$ Joules of energy in the form of neutrinos of all flavours at a rate of $10^{57}$ neutrinos per second. Marking the death of a massive star (mass >8-10 solar masses) and the birth of a neutron star or black hole, core collapse supernovae serve as laboratories for physics at extremes of density, temperature, and neutronization that cannot be produced in terrestrial laboratories. The kinetic energy and the rich mix of recently synthesized elements delivered into the interstellar medium by the ejecta of each supernova make core collapse supernovae a key link in our chain of origins from the Big Bang to the formation of life on Earth. Currently, the lack of data on neutrino-nucleus interactions limits our understanding of the mechanism by which core collapse supernovae explode our understanding of the resulting nucleosynthesis, and our ability to interpret the results of neutrino astronomy.

Recent studies [10] have demonstrated unequivocally that electron and neutrino capture reactions on nuclei play a major role in dictating the dynamics of stellar core collapse, which set the stage for all of post-bounce dynamics and the formation of the supernova shock wave. Comparisons of results using modern prescriptions for electron and neutrino capture on nuclei with earlier calculations demonstrate quantitative and qualitative changes in the launch radius of the supernova shock wave after stellar core bounce and in the density, temperature, and compositional structure of the stellar core.

Nucleosynthesis in core collapse supernovae falls into three basic categories: (1) explosive nucleosynthesis that occurs as the shock wave passes through the stellar layers and causes nuclear fusion through compression and heating of the material, (2) neutrino nucleosynthesis in the ejected layers that occurs as these layers are exposed to the intense neutrino flux emerging from the proto-neutron star, and (3) *r*-process nucleosynthesis that occurs in a neutrino-driven wind emanating from the proto-neutron star after the explosion is initiated. In all cases, the final elemental abundances produced and ejected are affected through nuclear transmutations by the neutrino-nucleus interactions that occur.

The ability to detect, understand, and ultimately use the detailed neutrino "light curve" from a future core collapse supernova in our galaxy is vital to better understanding supernovae and the use of supernova models, together with detailed astronomical observations, to constrain fundamental physics that is otherwise inaccessible in terrestrial experiments. To achieve this will require an accurate normalization of the neutrino flux in a supernova neutrino detector and knowledge of the cross sections and by-products of neutrino interactions in the detector material. From Deuterium to Lead, a number of nuclei have been proposed and, in some cases, used as supernova neutrino detector materials. In all cases, accurate neutrino-nucleus cross sections are essential. Only ν+Carbon has been measured to ~10% accuracy by LSND [11,12] and KARMEN [13]. With much lower accuracy (~40%) there are data for ν+Iron [14], ν+Iodine [15], and ν+Deuterium [16].

While it is impossible to experimentally measure all of the cross sections for the hundreds of weak interaction rates needed for realistic simulations of supernovae and supernova nucleosynthesis, a strategically chosen set of measurements will validate the fundamental nuclear structure and reaction models at the foundation of the hundreds of rate computations that are input for supernova models. Total neutrino-nucleus charged-current cross sections at low energy depend strongly on the charge number of the nucleus. For example the cross section in Lead is predicted to be 300 times that of Carbon. The charged-current reaction cross section induced by $\nu_e$ is thought to scale nearly as the square of the electron energy and is particularly sensitive to the detailed structure of the induced nuclear excitation spectrum. It is therefore important to either obtain the cross sections directly from experiment and/or calibrate theoretical models so that systematic uncertainties can be estimated.

In addition to being crucially important to astrophysics, measurements of neutrino-nucleus cross sections provide an opportunity to study issues of vital interest to nuclear structure related to the weak interaction. One of these involves understanding of the axial-vector response in nuclei. That response is often characterized by a modification of the axial vector coupling constant, in comparison to its value for free nucleons. For the Gamow-Teller operator that arises as a leading term from the low-energy (long wavelength) expansion of the weak axial current, one finds empirically that the effective axial-vector coupling constant is modified (quenched) in the nuclear medium. It is unknown to what extent other operators, of higher multipolarity, arising from the other terms in the expansion of the weak axial current are analogously modified. Using neutrinos from the SNS to probe medium energy strength distributions in neutrino-nucleus scattering would open the possibility to investigate this fundamental problem. Such measurements could be performed on target materials for which low-energy charge-exchange experiments are available in order to compare low-energy excitations.

Estimations shows that a ten ton active mass detector located at the twenty meters from the SNS target can detect thousands of neutrino-nucleus interaction per year, therefore enabling an extensive program of low energy neutrino-nucleus interactions. Details of the detector types that can be deployed can be found at [17], [18], [19]

## 4. Coherent Elastic Neutral Current Neutrino-Nucleus Scattering

Coherent elastic neutral current neutrino-nucleus scattering [20], [21] has never been observed. In this process, a neutrino of any flavor elastically scatters off a nucleus with low momentum transfer such that the nucleon wavefunction amplitudes are in phase and add coherently. The cross-section for such a process can be calculated exactly in the framework of the standard model, and for spin-zero nucleus, neglecting radiative corrections, is given by [22],

$$(d\sigma/dx)_{vA} = \frac{G_F^2 Q_W^2}{8\pi} F^2(2ME) M [2 - \frac{ME}{k^2}]$$

where $k$ is the incident neutrino energy, $E$ is nuclear recoil energy, $M$ is nuclear mass, $F$ is ground state elastic form factor $Q_w$ is the weak nuclear charge, and $G_f$ is the Fermi Constant. Condition for coherence is largely satisfied for neutrino energies up to ~50 MeV for medium mass nuclei.

For neutrino energies up to ~50 MeV, typical values of the total coherent elastic cross-section are in the range ~$10^{-39}$ cm$^2$, which is ten times large that inverse beta decay on protons and thousand times larger then elastic neutrino- electron scattering. In spite of its large cross-section this scattering is hard to observe due to the very low nuclear recoil energies, typically in ten's of keV.

Usually nuclear power plants with their large antineutrino fluxes have been considered as a prime location for the detection of coherent elastic neutral current neutrino-nucleus scattering [23],[24],[25]. Recently [26] it has been pointed that SNS has several key advantages relative to the nuclear plants for the detection of such process. Higher neutrino energies at SNS relative to the nuclear reactors give large recoil energies. This brings detection of such reaction within the reach of the current generation of typical dark mater detectors. The large duty cycle of SNS creates favorable background suppression conditions that are non-available at reactors. In addition signals from neutrons will show clear decay curves with the lifetime of the muon. This can clearly manifest detection of coherent scattering with unambiguous separation from beam related background. Details of the expected signal and rates can be found in [26] together with analysis of possible tests of the standard model that can be done via this process. Presently CLEAR collaboration is proceeding with proposal to deploy ~50kg noble liquid detector near SNS target station.

## 5. Tests Of The Standard Model

The electron spectrum from the reaction $^{12}C(v_e,e^-)^{12}N_{gs}$ can be used to derive the original $v_e$ spectrum from muon decay, taking into account $E_v = E_e + Q$ (17.8 MeV), the detector response function, and the $(E_{ve} - Q)^2$ dependence of the differential cross section. It is well known that a measurement of Michel parameters in muon decay is a direct test of the Standard Model and is a method to search for manifestations of new physics since these parameters are sensitive to the Lorentz structure of the Hamiltonian of weak interactions. The Michel parameters are related to the electron spectrum shape in muon decay. The neutrino spectrum from this decay can be described in terms of similar parameters.

In this manner the neutrino spectrum can provide complimentary information to the set of Michel parameters, increasing the accuracy and reliability in the search for new physics in the lepton sector.

As pointed out in [27], the shape of the $\nu_e$ spectrum from $\mu^+$ decay at rest is sensitive to scalar and tensor admixtures to pure V-A interactions due to the parameter $\omega_L$ which is analogous to the Michel parameter $\rho$, which determines the shape of the electron spectrum in muon decay. The $\nu_e$ spectrum can be written as:

$$dN_{\nu_e}/dx = \frac{G_F^2 m_\mu^5}{16\pi^3} Q_L^\nu (G_{0(x)} + G_1(x) + \omega_L G_2(x))$$

where $m_\mu$ is the muon mass, $x = 2E_\nu/m_\mu$ is the reduced neutrino energy, $Q_L^\nu$ is the probability for emission of a left-handed electron neutrino, $G_0$ describes pure V-A interactions, $G_1$ takes into account radiative corrections, and $\omega_L G_2$ includes effects of scalar and tensor components. $G_1$ is very small and can be neglected

In the Standard Model, $\omega_L$ is exactly zero. The KARMEN experiment [28] determined an upper limit for $\omega_L$: $\omega_L \leq 11\%$ (90% CL). Our calculations shows that at SNS due to the much higher neutrino flux even with modest 20 t liquid scintillator detector limit on $\omega_L$ can be lowered down to a few percent level.

## 6. Neutrino Oscillations and Lepton Flavour Number Violation in Muon Decay

Production of $\bar{\nu}_e$ is strongly suppressed at SNS relative to the other neutrino species. This is due to the very fast stopping and absorption of negative pions and muons at the SNS bulk mercury target. Calculations show that ratio of electron antineutrino to electron neutrino can be as low as a few per 10 000. It was point it out long time ago that due to this fact SNS is extremely attractive for the search for neutrino oscillations [29], [30]. Recently this interest has been revived by part of MiniBOONE collaboration [31]. A Kiloton class detector located at the distance of 60-70 meters can easily, in one year, unambiguously test the LSND claim.

Another interesting aspect that can be explored due to the low electron antineutrino rate is lepton number violation in muon decay. This has been clearly demonstrated in the neutrino sector in the form of neutrino flavor oscillations. Many extensions to the standard model, e.g. grand unified theories [32] and left-right symmetric models [33], violate lepton flavor number in exotic decays of the muon, e.g. $\mu^+ \to e^+ + \bar{\nu}_e + \nu_\mu$, which may be present at branching ratios as high as $10^{-4}$. [34]. SNS provides a special opportunity to search for lepton flavor number violating processes due to the small flux of electron antineutrinos and large fluxes of electron neutrinos produced. Electron antineutrinos can be easily observed by the inverse beta decay on proton reaction in liquid scintillator detector, located at the close proximity to the SNS target.

## 7. Conclusion
For this wealth of reasons, it is clear that a neutrino program at SNS would make an important contribution to the field of neutrino physics. Timely realization of this program will make significant impact on Astrophysics, Nuclear Physics and High-Energy physics.